# Deep-ultraviolet ptychographic pocket-scope (DART): mesoscale lensless molecular imaging with label-free spectroscopic contrast

Ruihai Wang[1,&], Qianhao Zhao[1,&,*], Julia Quinn[2], Liming Yang[1], Yuhui Zhu[1], Feifei Huang[1], Chengfei Guo[1], Tianbo Wang[1], Pengming Song[1], Michael Murphy[2], Thanh D. Nguyen[1,6], Andrew Maiden[3,4], Francisco E. Robles[5], and Guoan Zheng[1,6,*]

[1]Department of Biomedical Engineering, University of Connecticut, Storrs, CT 06269, USA
[2]Immunopathology Laboratory, University of Connecticut Health Center, Farmington, CT 06030, USA
[3]Department of Electronic and Electrical Engineering, University of Sheffield, Sheffield, SYK S1 3JD, UK
[4]Diamond Light Source, Harwell, Oxfordshire OX11 0DE, UK
[5]Wallace H. Coulter Department of Biomedical Engineering, Georgia Institute of Technology and Emory University, Atlanta, GA 30332, USA
[6]Center for Biomedical and Bioengineering Innovation, University of Connecticut, Storrs, CT 06269, USA
[&]These authors contributed equally
*Corresponding author: qianhao.zhao@uconn.edu or guoan.zheng@uconn.edu

**Abstract:** The mesoscale characterization of biological specimens has traditionally required compromises between resolution, field-of-view, depth-of-field, and molecular specificity, with most approaches relying on external labels. Here we present the Deep-ultrAviolet ptychogRaphic pockeT-scope (DART), a handheld platform that transforms label-free molecular imaging through intrinsic deep-ultraviolet spectroscopic contrast. By leveraging biomolecules' natural absorption fingerprints and combining them with lensless ptychographic microscopy, DART resolves down to 308-nm linewidths across centimeter-scale areas while maintaining millimeter-scale depth-of-field. The system's virtual error-bin methodology effectively eliminates artifacts from limited temporal coherence and other optical imperfections, enabling high-fidelity molecular imaging without lenses. Through differential spectroscopic imaging at deep-ultraviolet wavelengths, DART quantitatively maps nucleic acid and protein distributions with femtogram sensitivity, providing an intrinsic basis for explainable virtual staining. We demonstrate DART's capabilities through molecular imaging of tissue sections, cytopathology specimens, blood cells, and neural populations, revealing detailed molecular contrast without external labels. The combination of high-resolution molecular mapping and broad mesoscale imaging in a portable platform opens new possibilities from rapid clinical diagnostics, tissue analysis, to biological characterization in space exploration.

## 1 Introduction

The ability to visualize and analyze bio-specimens at the molecular level is fundamental to advancing life sciences and improving healthcare. Traditional imaging techniques have long relied on exogenous labels for contrast, often altering the natural state of bio-specimens and requiring lengthy staining procedures[1-3]. This limitation becomes critical in scenarios requiring rapid pathology diagnosis, such as fine-needle aspiration (FNA) cytology for cancer screening or intraoperative margin assessment, where conventional staining methods may delay crucial treatment decisions[4,5]. Label-free imaging methods have emerged to address these challenges without the need for stains[6-8]. However, simultaneously achieving high resolution, large field-of-view, extended depth-of-field, and molecular specificity remains difficult, particularly at the mesoscale where cellular and tissue-level processes intersect[9]. An ideal solution would involve quantitative measurement of molecular masses, providing a clear correlation between observed contrast and molecular composition for an explainable and trustworthy analysis.

To meet the challenges faced by traditional label-free imaging approaches, there is a growing interest in exploring different spectral regions that can provide intrinsic contrast based on the natural properties of biomolecules. The deep-ultraviolet (DUV) spectrum presents a largely untapped resource for label-free molecular imaging[10,11]. In this region, biomolecules such as proteins, nucleic acids, and lipids exhibit strong, distinctive absorption characteristics, offering potentials for intrinsic molecular contrast[10-17]. However, conventional cameras and lenses are incompatible with DUV wavelengths, necessitating specialized lenses that are prone to both geometrical and chromatic aberrations. In addition, lens-based approaches generally struggle to balance resolution, depth-of-field, and field-of-view for mesoscale imaging.

Here we introduce the Deep-ultrAviolet ptychogRaphic pockeT-scope (DART), a handheld lensless platform that transforms molecular imaging through deep-ultraviolet spectroscopic contrast. DART leverages the intrinsic absorption characteristics of biomolecules in the deep-ultraviolet spectrum, where proteins and nucleic acids exhibit distinct spectral signatures. Through differential spectroscopic imaging, DART quantitatively maps molecular mass distributions with femtogram sensitivity, enabling explainable virtual staining without external labels. By employing lensless ptychographic reconstruction, DART achieves 308-nm resolution across centimeter-scale areas with a depth-of-field extending to millimeters -- capabilities unmatched by conventional microscopy. The system's virtual error-bin methodology effectively eliminates artifacts from limited temporal coherence and optical imperfections, ensuring high-fidelity molecular contrast.

The DART platform addresses several fundamental challenges in label-free molecular imaging. In cytopathology, DART's ability to quantitatively map nucleic acid and protein mass distributions enables rapid tissue analysis with molecular specificity. Unlike traditional staining methods that require lengthy preparation or black-box deep learning approaches that lack interpretability, DART provides explainable virtual staining through direct measurement of molecular content. In tissue analysis, DART's large field-of-view and extended depth-of-field enable comprehensive examination of entire tissue sections without mechanical refocusing, providing detailed molecular maps across large sample areas. For neuroscience research, DART facilitates quantitative mapping of neural populations through their intrinsic molecular signatures, revealing brain architecture without perturbing native molecular states. The system's compact and handheld design further extends these capabilities to space-constrained and other challenging environments, from resource-limited field settings to deployment in space missions, where it could enable studies in personalized medicine and space biology. By combining mesoscale imaging capabilities with quantitative molecular contrast, DART establishes a new paradigm for label-free biological characterization, opening possibilities from clinical laboratories to space exploration.

## 2 Results
### 2.1 Design and overview of the DART system
Figure 1 provides an overview of the DART system design, operation, and capabilities. As shown in Fig. 1a, the DART device integrates three illumination sources -- two DUV LEDs at 266 nm and 280 nm for molecular imaging, and a 405-nm laser diode for complementary phase imaging. In the current implementation, these light sources are activated sequentially to capture wavelength-specific information, though the system architecture could support simultaneous multi-wavelength illumination for multiplexed imaging[18, 19]. The reconstruction process incorporates virtual states to isolate and remove artifacts arising from optical imperfections, ensuring high-fidelity image recovery.

The optical layout in Fig. 1b reveals the complete experimental configuration with a two-panel view. The top-view panel illustrates how UV-enhanced mirrors create a folded beam path of approximately 20 cm through multiple reflections, transforming the diverging light from point sources into illumination that covers the entire imaging area. This folded-path design achieves uniform sample illumination without beam-forming optics, maintaining system simplicity for handheld operation. The main panel shows the spatial arrangement of all components: the three light sources positioned at the top, the sample placement area, and the coded sensor assembly at the bottom. Supplementary Fig. S1 provides comprehensive documentation of the DART setup with detailed component labelling and assembly sequence, as well as experimental pipeline. The sensor assembly consists of a modified CMOS sensor where the original microlenses have been etched away to enhance DUV sensitivity -- a modification that doubles the sensor's response at DUV wavelengths (Supplementary Fig. S2). A disorder-engineered coded surface is attached directly to this modified sensor, containing random phase and amplitude modulating elements that convert high-angle scattered light into detectable signals, effectively increasing numerical aperture and resolution beyond conventional sensor limits[20]. The critical distances are clearly labeled in Fig. 1b: samples are positioned 0.2-2 mm from the coded surface (optimized post-measurement via digital back propagation), while the coded surface maintains a fixed 0.84 mm separation from the pixel array. These short propagation distances ensure operation in the Fresnel diffraction regime[21, 22], enabling both the compact design necessary for portability and the large field of view with millimeter-scale depth of field that distinguishes DART from conventional microscopy approaches.

During DART's operation, illumination and sample remain stationary while the entire sensor assembly, including the coded surface, shifts in a grid pattern with 1-3 µm randomized steps to capture a set of diffraction patterns. The sensor-shift actuator comprises magnets and coils with motion constraint rails, as shown in Supplementary Fig. S1, Note 1, and Videos S1-S2, with the design inspired by optical image stabilization technology used in smartphone photography. This sensor-shift approach, which contrasts with conventional ptychography, offers several practical advantages for a handheld device. It simplifies the mechanical design by eliminating alignment complexities associated with beam steering optics and accommodates multiple light sources without wavelength-specific scanning components. Compared with specimen translation, moving the sensor assembly enables in-situ imaging of live samples in culture dishes -- moving such samples can be challenging as the vibration of the liquid inside may introduce phase artifacts during data acquisition.

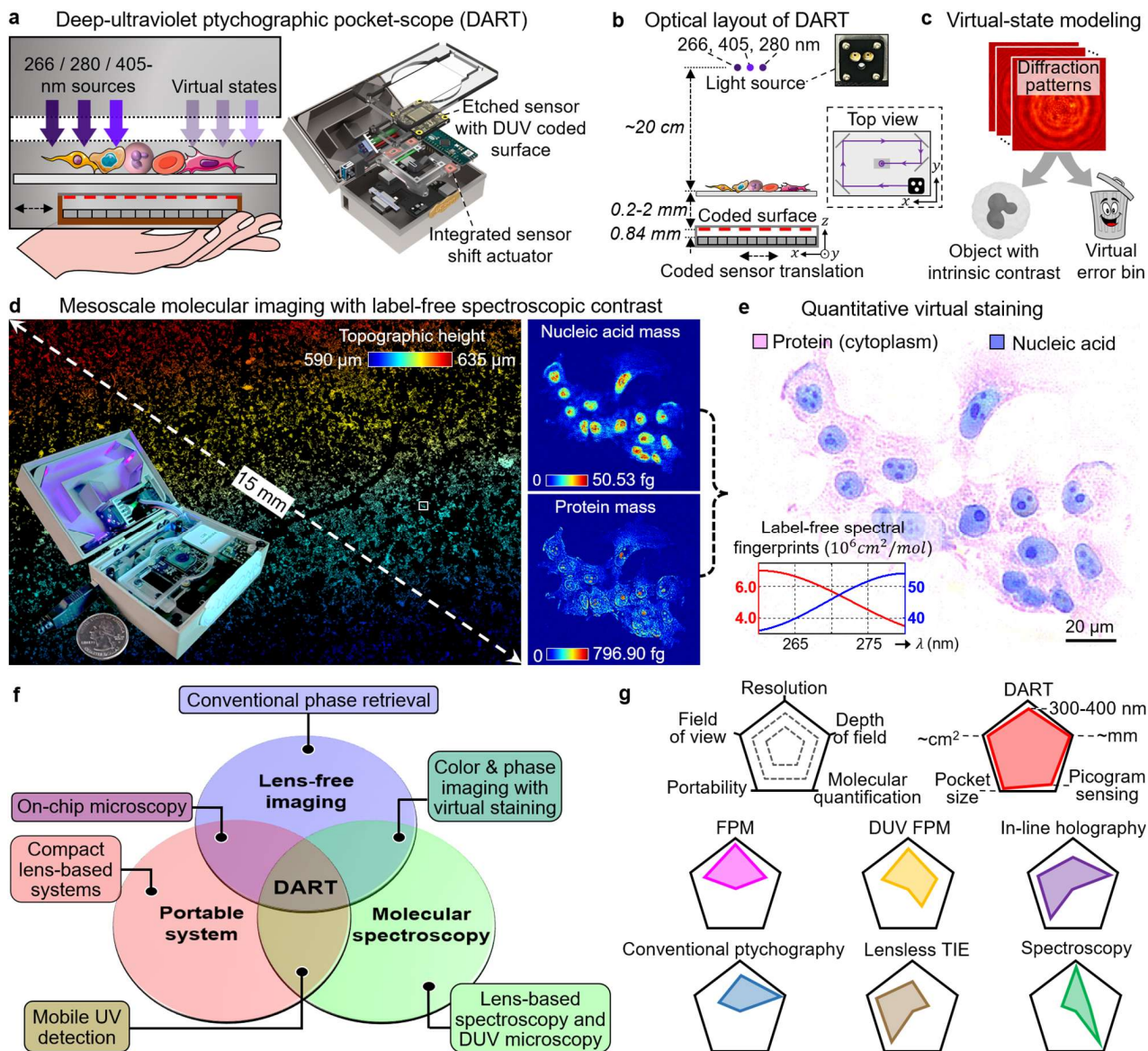

**Fig. 1| Overview of the DART system. a**, The DART design featuring three illumination sources that are sequentially activated for differential spectroscopic imaging. The compact system integrates these light sources with a sensor assembly for handheld operation, incorporating virtual state methodology for reconstruction. **b**, Optical layout of DART showing the folded beam path via mirrors (top-view panel), with light sources, sample, and coded sensor assembly positioned at the labeled distances. **c**, Virtual state enabled recovery methodology, separating the object information from errors in the error bin. **d**, DART acquires sub-cellular details over centimeter-scale areas with intrinsic molecular contrast. **e**, Quantitative virtual staining based on the recovered molecular mass distributions. The inset show distinct absorption characteristics of nucleic acids (blue) and proteins (red) at DUV regime. **f**, Three key technological domains in biomedical imaging: lens-free imaging[23-25], portable systems[26, 27], and molecular spectroscopy[3, 10-17]. While pairwise combinations have enabled various developments[22, 28-39], DART uniquely integrates all three domains. **g**, Performance comparison with common imaging approaches across five metrics. For detailed documentation of the setup, component, assembly instructions, and experimental pipeline, see Supplementary Fig. S1, Note 1, and Videos S1-S2.

Figure 1c shows the virtual state methodology employed in DART, which reduces artifacts caused by various factors including limited temporal coherence of DUV sources, multi-reflection fringes, and other system errors. As detailed in Supplementary Fig. S3 and Note 2, the reconstruction process consists of two paths: a primary path representing the true object states using an accurate forward imaging model, and a secondary path with a perturbed forward model to capture systematic errors. Unlike object modes, virtual error bins operate by deliberately introducing a perturbation to model the discrepancies between the ideal imaging model and real-world imperfections. In our implementation, this perturbation involves modifying the propagation distance parameter in the forward model by a factor '$a$' in Supplementary Fig. S3, creating a computational mismatch that effectively functions as a virtual error bin. This approach differs fundamentally from multi-layer ptychography methods[40, 41], which model sequential wave propagation through physical layers of 3D thick samples. Instead, our virtual error-bin states serve as computational constructs that isolate signal components inconsistent with the accurate object model, without representing physical sample layers. The incoherent summation of intensities from both paths allows systematic errors to be diverted into the virtual state, leaving the object reconstruction largely artifact-free.

The system's imaging capabilities are demonstrated in Fig. 1d-1e. Each wavelength's diffraction patterns undergo ptychographic reconstruction to recover the complex wavefield at the coded surface plane, which can then be digitally propagated back to any axial plane for refocusing. The topographic height map generated through this post-measurement refocusing is presented in the left panel of Fig. 1d. When using DUV wavelengths, the reconstructed amplitude images at 266 nm and 280 nm can be processed to map protein and nucleic acid distributions, as shown in the right panel of Fig. 1d. This quantitative molecular information forms the basis for explainable virtual staining in Fig. 1e, where recovered molecular masses are translated into familiar histological appearances. This capability offers an explainable alternative to black-box deep learning approaches for virtual staining[10, 14-17], providing direct correlation between measured molecular content and image contrast.

In Fig. 1f, we show that microscopy imaging encompasses three key technological domains: 1) lens-free imaging, which includes conventional phase retrieval methods such as in-line holography[42-47], multi-height/wavelength reconstructions[36, 37, 39, 48-50], and transport-of-intensity algorithms[51, 52]; 2) portable systems that prioritize compact and accessible designs[26, 27]; and 3) molecular spectroscopy featuring lens-based spectroscopy and DUV microscopy for biomolecular analysis[3, 10-17]. The convergence of these domains has produced various hybrid approaches. The combination of lens-free imaging with molecular spectroscopy has enabled color and phase imaging with virtual staining capabilities[28, 29], while the integration of portable systems with molecular spectroscopy has yielded mobile UV detection platforms[30-32]. The merger of lens-free imaging and portable systems has produced on-chip microscopy solutions[22, 33-39]. DART uniquely operates at the intersection of all three domains, synthesizing their capabilities into a single integrated platform.

Performance analysis across five critical metrics demonstrates DART's comprehensive advantages (Fig. 1g). Existing techniques show specific strengths but face inherent limitations. Conventional Fourier ptychographic microscopy (FPM)[53-55] achieves high resolution but lacks molecular contrast. DUV FPM[56] provides molecular contrast with its single-wavelength DUV LED array. However, its image quality and field of view suffer from chromatic and geometric aberrations of the UV objective lens system, resulting in a demonstrated field of view substantially smaller than visible light. In-line holography[42-47] excels in depth of field and portability but provides insufficient data diversity for robust phase recovery, particularly challenging with DUV LED sources due to their limited spatial and temporal coherence. Conventional ptychography[57-60] achieves high resolution but is constrained in field of view due to its spatially-confined probe beam illumination, while lensless TIE[51, 52] enables phase imaging over large areas but cannot differentiate molecular species. Lens-based spectroscopy[3, 10-17] provides molecular specificity but with restricted field of view and depth. DART uniquely combines the strengths: up to 308-nm resolution, centimeter-scale field of view, millimeter-scale depth of field, and picogram-level molecular quantification in a pocket-sized platform.

**2.2 Resolution characterization**

Figure 2 presents the resolution characterization of DART. Figure 2a shows a raw diffraction pattern captured at 280 nm, with a magnified view highlighting the pixelated measurement. The pre-calibrated deep-ultraviolet coded surface on the modified image sensor is illustrated in Fig. 2b, where amplitude is represented by grayscale values and phase is encoded by color hues. Proper characterization of the partially coherent DUV LED source is essential for improving image resolution, as demonstrated in Figs. 2c and 2d. Without accounting for the partial spatial coherence of the DUV LED source, the reconstruction shows reduced resolution in Fig. 2c. In contrast, by modeling the shape of the light source in Fig. 2d, DART successfully resolves features down to 435 nm on the resolution target. This capability is further illustrated in the imaging of label-free monocytes in Figs. 2e-2f, where source characterization enables clear visualization of sub-cellular structures typically challenging to resolve in a label-free manner. A comparison with a conventional microscope is presented in Fig. 2g, where a 20×, 0.75-NA objective lens captures an image of the same monocytes in the blood smear. Unlike the microscope image, DART's reconstruction reveals intrinsic molecular contrast without labels, highlighting its distinct advantage in label-free biological imaging.

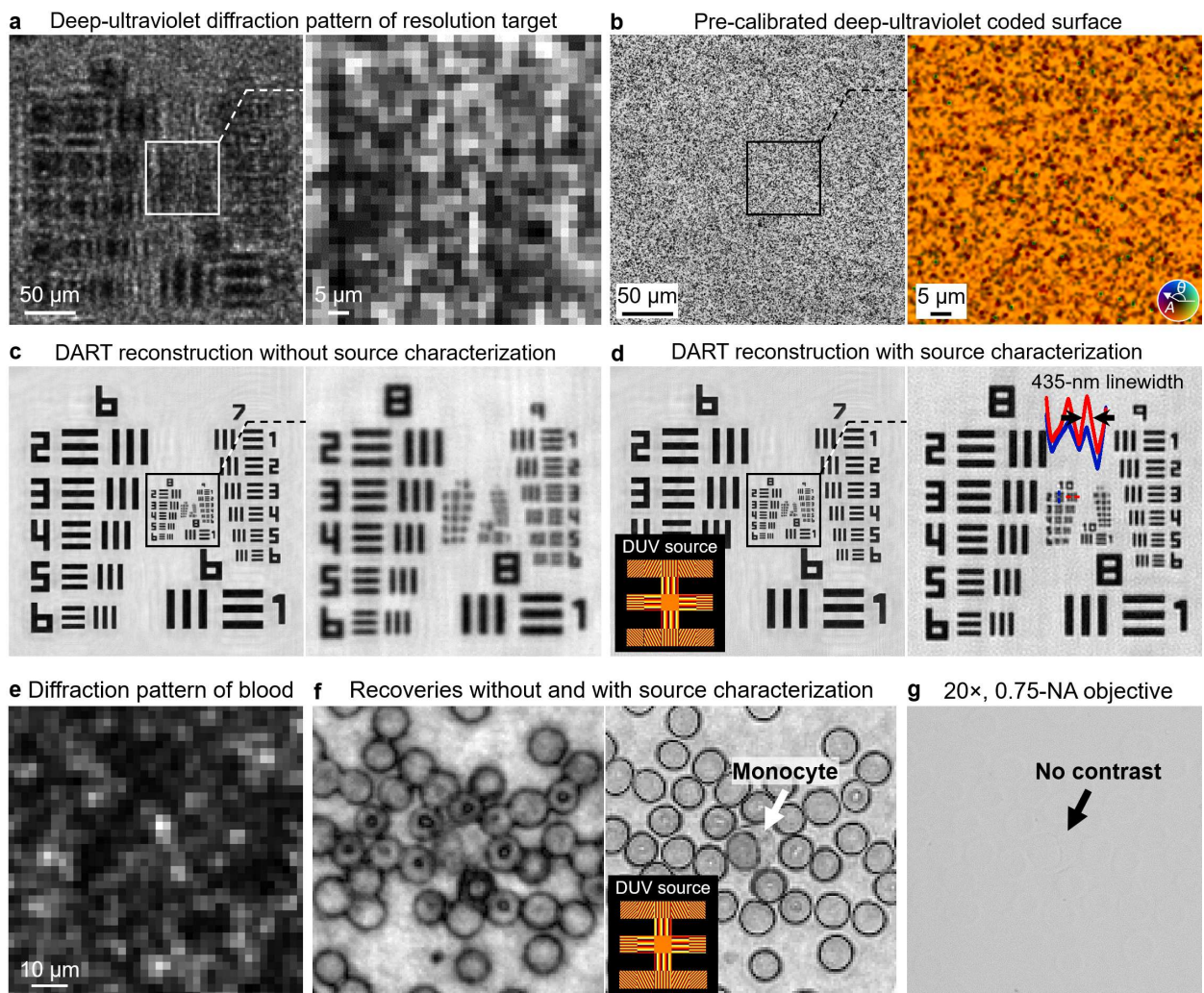

**Fig. 2| High-resolution diffraction imaging and source characterization with DART**. **a**, Raw diffraction pattern captured by DART. **b**, Pre-calibrated deep-ultraviolet coded surface on the modified image sensor, where amplitude is represented by grayscale values and phase is encoded by color hues. **c**, DART reconstruction without source characterization, where the partial spatial coherence of the DUV light source is not accounted for, leading to reduced image resolution. **d**, By modeling the shape of the DUV light source in reconstruction, DART successfully resolves the 435-nm linewidth on the resolution target. By using the 405-nm laser diode source, DART can further resolve the 308-nm linewidth on the resolution target (Supplementary Fig. S5). **e**, Raw diffraction pattern of blood cells captured by DART. **f**, DART reconstructions without (left) and with (right) source characterization. Source characterization enables the clear visualization of label-free sub-cellular structures of monocyte. **g**, Image of the same unstained blood smear captured using a conventional 20×, 0.75-NA objective lens. The low contrast in this image highlights the advantages of DART in revealing intrinsic molecular contrast in biological specimens without the need for external labels.

Supplementary Fig. S4 further emphasizes the importance of source characterization in DART imaging. It presents raw diffraction images and reconstructions of unstained HEK-293 cell cultures and blood smears, both with and without source characterization. The reconstructions with proper source characterization reveal sub-cellular structures in the cell cultures and individual cells. Supplementary Fig. S5 extends the resolution characterization across DART's three wavelengths. At the DUV wavelengths of 266 nm and 280 nm, DART resolves the 435-nm linewidth on the resolution target. Notably, utilizing the laser diode at 405 nm, DART resolves the 308-nm linewidth. The better resolution achieved at the 405-nm wavelength is attributed to the superior spatial and temporal coherence of the source. This observation also highlights the potential for further improving DUV resolution by employing a more precise model to account for the coherence properties of DUV light sources.

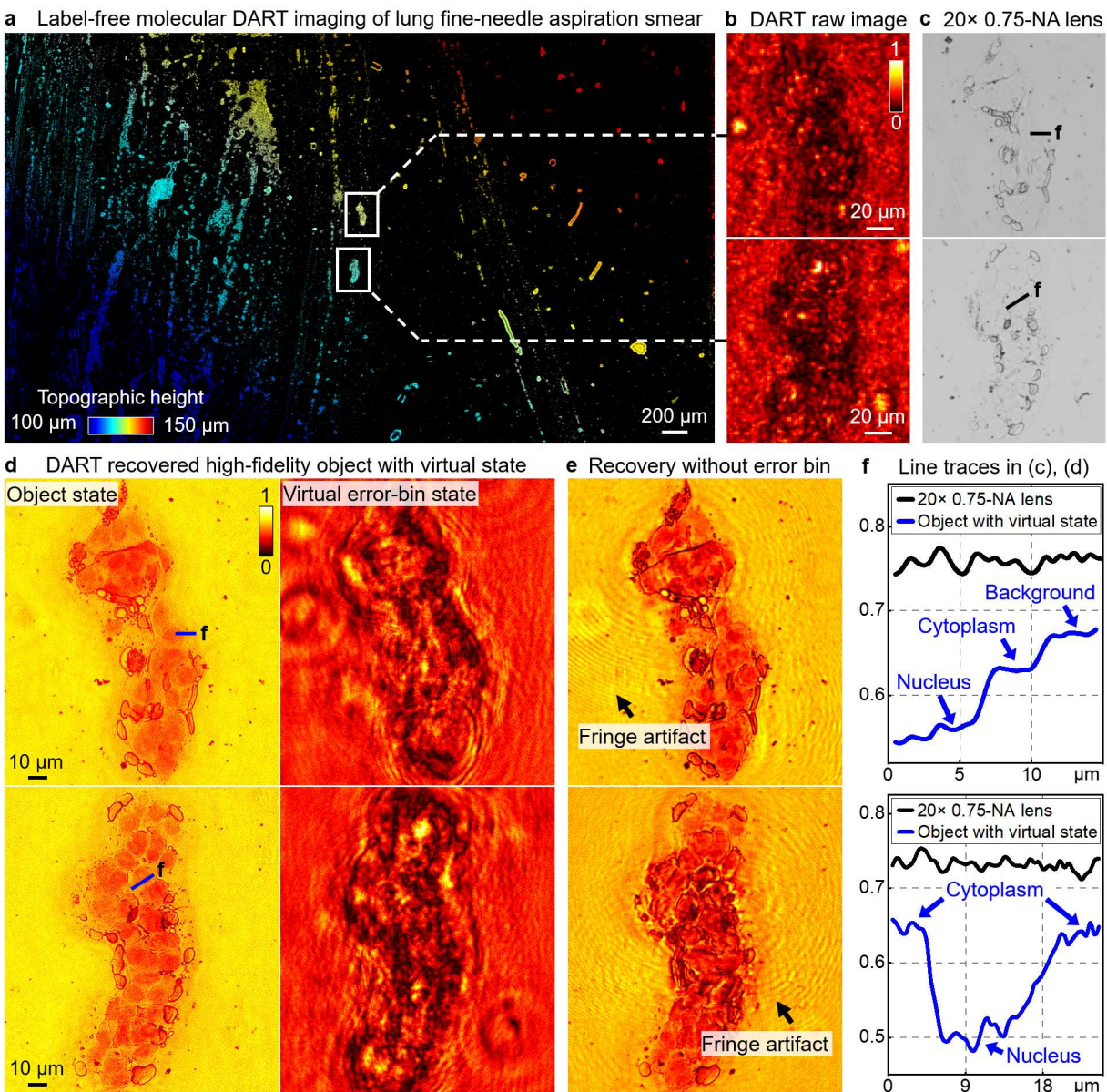

**Fig. 3 | High-fidelity mesoscale DART imaging of unstained cytopathology sample with virtual state correction. a**, Topological height map of a label-free lung FNA smear generated post-measurement, showing topological height variation across a large imaging area. **b**, Raw diffraction measurements. **c**, Images captured using a conventional 20×, 0.75-NA objective lens, showing no contrast across the marked green lines. **d**, DART reconstruction with virtual state correction. Intensity traces through the marked blue lines reveal cytoplasm and nucleus with intrinsic molecular absorption, demonstrating DART's ability to image label-free FNA smears for applications in cytological analysis. The virtual error-bin state isolates and removes artefacts caused by the limited temporal coherence of DUV sources, multi-reflection fringes, and other noises, ensuring high-fidelity and accurate reconstruction. **e**, DART reconstruction without virtual state correction reveals residual artifacts and reduced image quality, highlighting the impact of uncorrected system errors on image fidelity. **f**, Line traces comparing object reconstruction with virtual state correction (blue) and conventional light microscope (black). Also refer to the DART reconstruction comparison with and without the virtual state in Supplementary Video S3.

## 2.3 Mesoscale DUV imaging with intrinsic molecular contrast

DART's ability to perform imaging with intrinsic molecular contrast is demonstrated with different label-free biospecimens. Figure 3 illustrates DART's capabilities in imaging an unstained lung FNA smear. Figure 3a shows the focus map of the sample generated post-measurement, revealing topographical height variations across a large imaging area. In DART, the distance between the sample and the coded surface is at the millimeter scale. We employ the Brenner gradient metric to determine the optimal refocusing distance[61]. Figure 3b shows the raw diffraction measurements of two regions of interest, and Fig. 3c shows the conventional light microscope images captured using a 20×, 0.75-NA objective lens. In Figs. 3d-3e and Supplementary Video S3, we compare DART reconstructions with and without virtual state correction. With virtual state correction (Fig. 3d), intensity traces through the marked lines reveal the cytoplasm and nucleus with intrinsic molecular absorption, demonstrating DART's ability to image unstained FNA smears for cytological analysis. The virtual error-bin state effectively isolates and removes artifacts, as shown in the right panel of Fig. 3d. In contrast, Fig. 3e shows the DART reconstruction without virtual state correction, revealing residual artifacts and reduced image quality. In Fig. 3f, we provide line traces comparing DART reconstructions to conventional light microscopy, highlighting DART's intrinsic contrast in label-free samples.

Supplementary Fig. S6 extends the comparison using the same lung FNA smear sample, contrasting DART with conventional lens-based brightfield and fluorescence methods. As shown in Supplementary Fig. S6, DART resolves sub-cellular structures of the cytoplasm and nuclei without external labels or stains. DAPI fluorescence imaging confirms the structural features observed in the DART images. Additionally, line tracing of different small features demonstrates that DART's performance is in good agreement with images captured using a 100×, 1.25-NA oil-immersion objective.

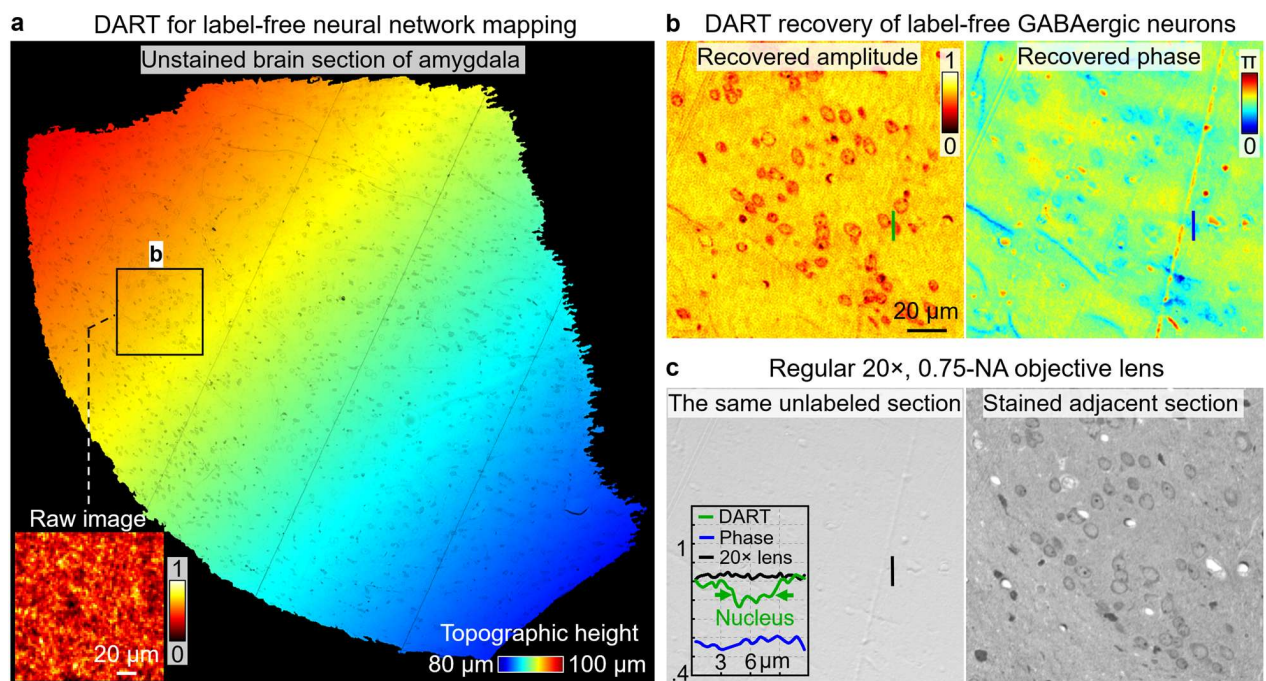

**Fig. 4 | DART imaging of unlabelled brain section for label-free neural mapping with intrinsic molecular contrast. a**, DART imaging of an unlabelled brain section of amygdala, with the inset showing the raw diffraction pattern. **b**, DART recovery of GABAergic neurons. The recovered DART amplitude (left) provides clear contrast, resolving GABAergic neurons. In contrast, the recovered phase image (right) offers little contrast, making it difficult to distinguish individual cells. **c**, Imaging the same region with a conventional brightfield microscope shows no contrast (left). The adjacent stained section (right) confirms the presence of GABAergic neurons, which DART resolves in a label-free manner. The inset of the left panel displays line profiles comparing the intrinsic contrast provided by DART with limited contrast of phase and conventional microscope. The intrinsic contrast makes it a new tool for studying neuronal connectivity without the need for external labeling.

Figure 4 demonstrates DART's ability to image unstained mouse brain sections, allowing neural cell mapping with intrinsic contrast. Figure 4a shows DART imaging of an unlabelled brain section of amygdala, with the inset showing the raw diffraction pattern of a region of interest. Figure 4b presents DART's recovery of GABAergic neurons, with the recovered DART amplitude (left) providing clear contrast and resolving individual neurons. Interestingly, the recovered phase image (right) offers little contrast, highlighting the unique advantage of DART's molecular contrast in label-free imaging. Figure 4c shows the images captured using a conventional light microscope with 20×, 0.75-NA objective lens. The left panel of Fig. 4c shows no contrast on the same unlabelled brain section, while the right panel shows the image of the adjacent stained section, confirming the presence of GABAergic neurons that DART resolves in a label-free manner. The ability to visualize neuronal populations without labels could accelerate studies on brain architecture, neuronal connectivity, and how these are altered in various neurological disorders.

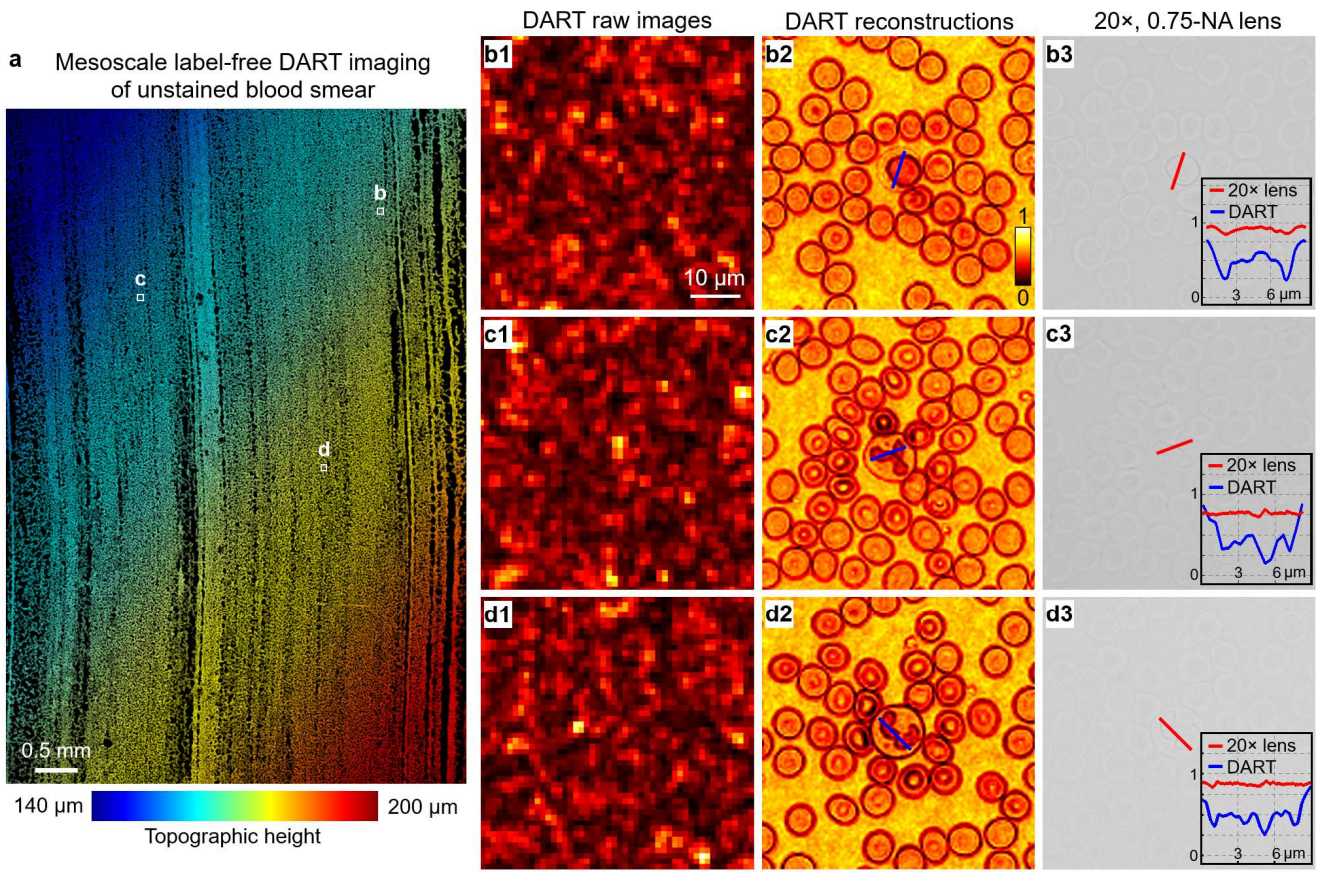

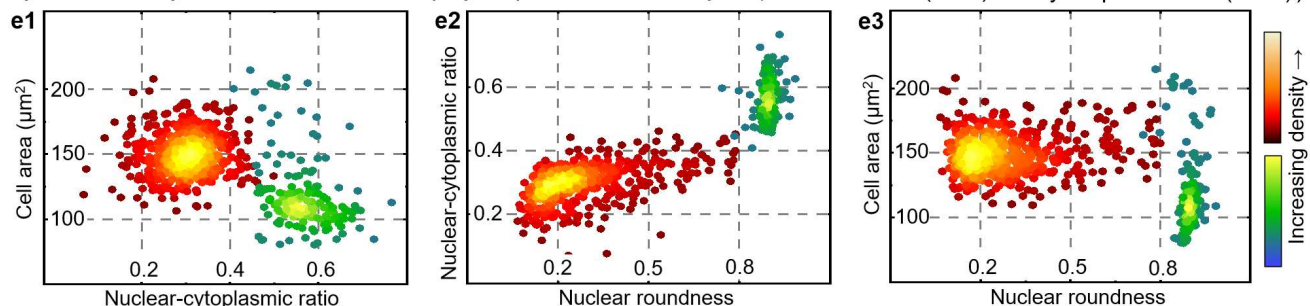

**Fig. 5 | Mesoscale label-free DART imaging of unstained blood cells. a**, Mesoscale DART imaging of a label-free blood smear sample, with topographic height variations encoded in color. **b1-d1**, DART raw images from three distinct regions. **b2-d2**, DART reconstructions, revealing individual leukocytes. **b3-d3**, Images with 20×, 0.75-NA lens show no discernible contrast. The insets show the features recovered by DART (blue) compared to the the regular 20× lens (red). **e1-e3**, Cytometric analysis of mononuclear (MNL, green) and polymorphonuclear (PMN, red) leukocytes based on DART imaging. **e1**, Cell area versus nuclear-cytoplasmic ratio distribution. **e2**, Nuclear-cytoplasmic ratio versus nuclear roundness plot. **e3**, Cell area versus nuclear roundness distribution. The clear separation between MNL and PMN populations demonstrates DART's potential for rapid, label-free blood cell differential analysis without external labels.

Figure 5 demonstrates DART's ability to perform label-free imaging of blood cells across centimeter-scale areas, revealing detailed cellular features without conventional staining procedures. As shown in Fig. 5a, DART captures high-resolution with topographic information across a large field-of-view. Figures 5b1-5d1 shows the captured raw images of selected regions of interest. Using intrinsic DUV molecular contrast, DART reconstructs high-resolution images of individual leukocytes (Figs. 5b2-5d2), clearly resolving cellular features that remain invisible under conventional brightfield microscopy (Figs. 5b3-5d3). The system's automated cytometric analysis reveals distinct clustering of leukocyte populations based on multiple quantitative parameters, including nuclear-cytoplasmic ratios (Fig. 5e1), nuclear roundness (Fig. 5e2), and cellular area (Fig. 5e3). This label-free differentiation enables clear separation between mononuclear and polymorphonuclear cells. The relative proportions of these leukocyte populations serve as important diagnostic indicators[62], as shifts in the mononuclear-to-polymorphonuclear ratio often suggest specific pathological conditions—elevated lymphocyte counts typically indicate viral infections, while increased neutrophil populations suggest bacterial infections. DART's ability to provide this morphological information without sample preparation or staining represents an advance for rapid hematological analysis. The system's compact size and label-free operation make it particularly valuable for point-of-care facilities, where rapid blood analysis can guide critical treatment decisions.

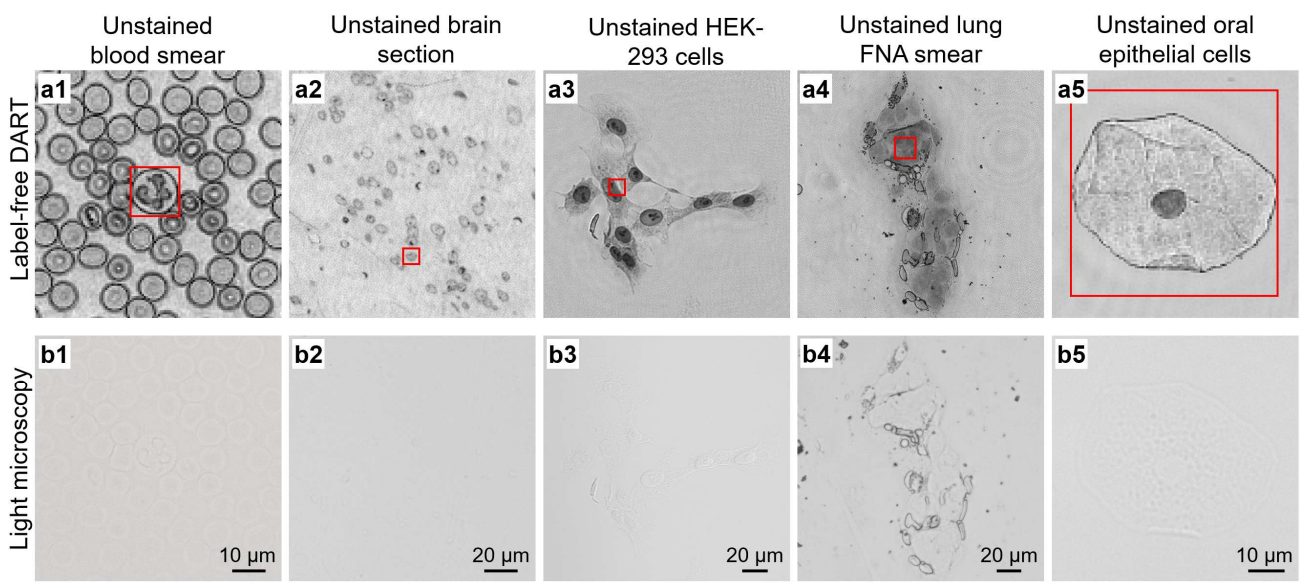

| | Blood smear | Neural cells in brain section | HEK-293 cells | Lung FNA smear | Oral epithelial cells |
|---|---|---|---|---|---|
| Conventional lens-based microscopy | 0.0485 | 0.0420 | 0.0728 | 0.0631 | 0.0795 |
| Label-free DART | 0.9255 | 0.3868 | 0.7627 | 0.6922 | 0.9205 |
| Improvement with DART | 19.0× | 9.2× | 10.4× | 10.9× | 11.5× |

**Fig. 6 | Contrast comparison between lens-based light microscopy and DART across various unstained bio-samples. a**, Label-free DART imaging of various unstained biological samples, including blood smear **a1**, brain section **a2**, HEK-293 cells **a3**, lung FNA smear **a4**, and oral epithelial cells **a5**. DART reconstructions provide high contrast and detailed morphological information for each sample. The red boxes highlight the regions of interest for contrast analysis. **b**, Corresponding images captured by conventional light microscopy using a 20×, 0.75-NA objective lens of the same samples as in **a**. The unstained specimens show little contrast, with minimal cellular features visible. **c**, Quantitative contrast comparison between lens-based light microscopy and DART for different bio-samples. The table provides the contrast values for each imaging method, demonstrating the significant improvement in contrast offered by DART. For all samples, DART achieves an increase in contrast ranging from 9.2× to 19.0× compared to conventional lens-based microscopy.

In Supplementary Figs. S7 and S8, we demonstrate the effectiveness of virtual state correction for imaging unstained cell culture samples and pathology slides. By removing artifacts and error sources, DART improves image quality and ensures that the intrinsic molecular contrast is well-represented. The system's capability to reveal intrinsic contrast without external labels offers new possibilities for rapid tissue analysis in research and clinical settings.

Figure 6 provides a comprehensive quantitative and qualitative comparison between DART and conventional lens-based microscopy across five different unstained bio-samples. Figures 6a1-a5 show DART's label-free imaging capabilities for blood smear, brain section, HEK-293 cells, lung FNA smear, and oral epithelial cells, respectively, revealing detailed cellular structures and molecular contrast. In contrast, conventional brightfield microscopy images of the same samples (Figs. 6b1-b5) show minimal visible contrast, making it difficult to distinguish cellular features without external staining. The quantitative contrast analysis (Fig. 6c) demonstrates DART's superior performance across all sample types: blood smear imaging shows a 19.0× improvement (DART: 0.9255 vs. conventional: 0.0485), neural cells in brain sections exhibit a 9.2× enhancement (0.3868 vs. 0.0420), HEK-293 cells display a 10.4× increase (0.7627 vs. 0.0728), lung FNA smear demonstrates a 10.9× improvement (0.6922 vs. 0.0631), and oral epithelial cells show an 11.5× enhancement (0.9205 vs. 0.0795). These dramatic improvements in contrast for label-free imaging hold particular promise for rapid diagnostic applications, where traditional staining procedures can introduce delays in clinical decision-making. The consistent performance across diverse sample types—from liquid blood smears to solid tissue sections—demonstrates DART's versatility and potential impact across multiple clinical and research applications.

## 2.4 Spectroscopic molecular fingerprints for quantitative virtual staining

DART's capabilities extend beyond imaging at a single DUV wavelength. Through differential spectroscopic imaging, it can recover nucleic acid mass and protein mass distributions, which can be used to perform quantitative virtual staining. Figure 7 demonstrates this capability using HEK 293 cell culture samples, where Figs. 7a and 7b show DART's recovered images at 266 nm and 280 nm. By exploiting the absorption characteristics of biomolecules at these two DUV wavelengths[10-12, 16], DART maps nucleic acid and protein mass distributions, as shown in Fig. 7b, with color bars indicating the mass distribution in femtograms per pixel. This quantitative information forms the basis for explainable and trustworthy virtual staining, as illustrated in Fig. 7c. The left panel shows cytoplasmic (purple) and nucleic acid (blue) distributions, while the right panel presents an H&E-like virtual stain, derived directly from the quantitative maps (Methods). Unlike deep-learning-based virtual staining, DART's approach is grounded in measured molecular content, offering an inherently explainable basis for image contrast.

Figure 7d provides a comparative image using a conventional 20×, 0.75-NA objective, showing no contrast of the label-free sample. In Fig. 7e, we compare line traces of nucleic acid mass, protein mass, and brightfield intensity, clearly illustrating DART's superior molecular contrast. DART's ability to quantitatively recover molecular content is further validated using unstained epithelial cells from a buccal smear in Figs. 7f-7i. Figure 7f presents nucleic acid and protein mass maps of the unstained epithelial cells. Figure 7g shows the corresponding quantitative virtual stain, effectively differentiating between cellular components. This is contrasted with a conventional microscopy image in Fig. 7h, which lacks any visible contrast. The line traces in Fig. 7i confirm the superior contrast provided by DART compared to traditional light microscopy.

Figure 8 demonstrates label-free molecular imaging of pathology sections. Figure 8a shows the topographic height map generated post-measurement. Figure 8b shows the raw measurements and the recovered DART images at the two DUV wavelengths, which provide context for the quantitative molecular mapping, while Figs. 8c and 8d compare the label-free recoveries of nucleic acid and protein mass distributions with the recovered phase image. Notably, cell nuclei are clearly identifiable through nucleic acid content in the mass distribution maps but are not distinguishable in the phase image, highlighting DART's advantage over traditional phase imaging techniques. Figure 8e shows the virtually stained images based on the recovered molecular distributions (Methods). In contrast, a comparative image taken with a regular 20×, 0.75-NA objective in Fig. 8f shows no contrast in this sample.

Supplementary Fig. S9 further illustrates DART's ability to identify cellular structures in unstained samples compared to conventional brightfield and phase microscopy. While these traditional methods struggle to provide clear contrast, DART's nucleic acid and protein mass maps reveal detailed cellular structures. This enables the production of virtual H&E stains that closely resemble traditional staining methods. Supplementary Fig. S10 further confirms this resemblance by directly comparing DART's virtual H&E stain with a traditional color micrograph from an adjacent stained section, demonstrating the virtual stain's fidelity in representing cellular structures according to their molecular makeup. DART's approach to virtual staining provides a direct link between measured molecular content and resulting image contrast, ensuring explainability. The label-free nature of the technique preserves the native state of the sample, while the spectroscopic approach allows for potential multiplexed imaging of various molecular species simultaneously[15].

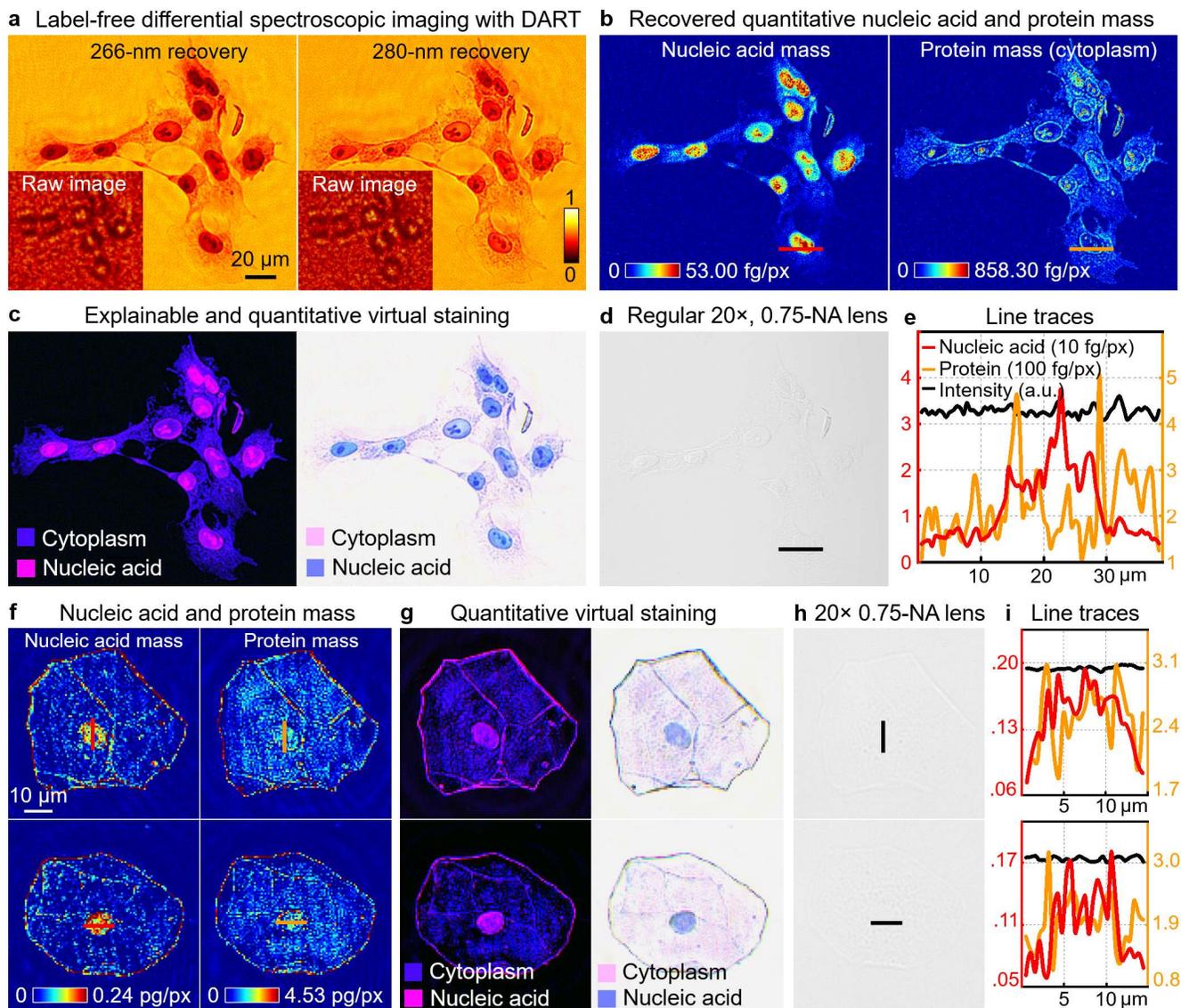

**Fig. 7 | Quantitative recovery of nucleic acid and protein mass for label-free, trustworthy virtual staining. a**, Differential spectroscopic imaging of label-free HEK-293 cells at two DUV wavelengths: 266 nm (left) and 280 nm (right). The recovered images highlight the intrinsic molecular contrast with the raw diffraction measurements shown in the insets. **b**, Recovered quantitative nucleic acid and protein mass maps using differential DUV imaging, with the color bars indicating quantitative mass distribution in femtograms per pixel (fg/px). **c**, Explainable and quantitative virtual staining based on the recovered nucleic acid and protein mass. The image on the left shows cytoplasm (purple) and nucleic acid (blue) distributions, while the right panel presents a H&E stain appearance based on the two quantitative distribution maps. **d**, Regular 20×, 0.75-NA lens image shows no contrast in the label-free sample. **e**, Line traces comparing nucleic acid mass (red), protein mass (yellow), and regular brightfield intensity (black) along the line profile in (**b**) and (**d**). **f**, Nucleic acid and protein mass maps for label-free epithelial cells from a buccal smear. **g**, Quantitative virtual staining based on the two quantitative maps in (**f**). **h**, Regular 20×, 0.75-NA lens image. **i**, Line traces corresponding to the nucleic acid and protein mass in (**f**) and (**h**).

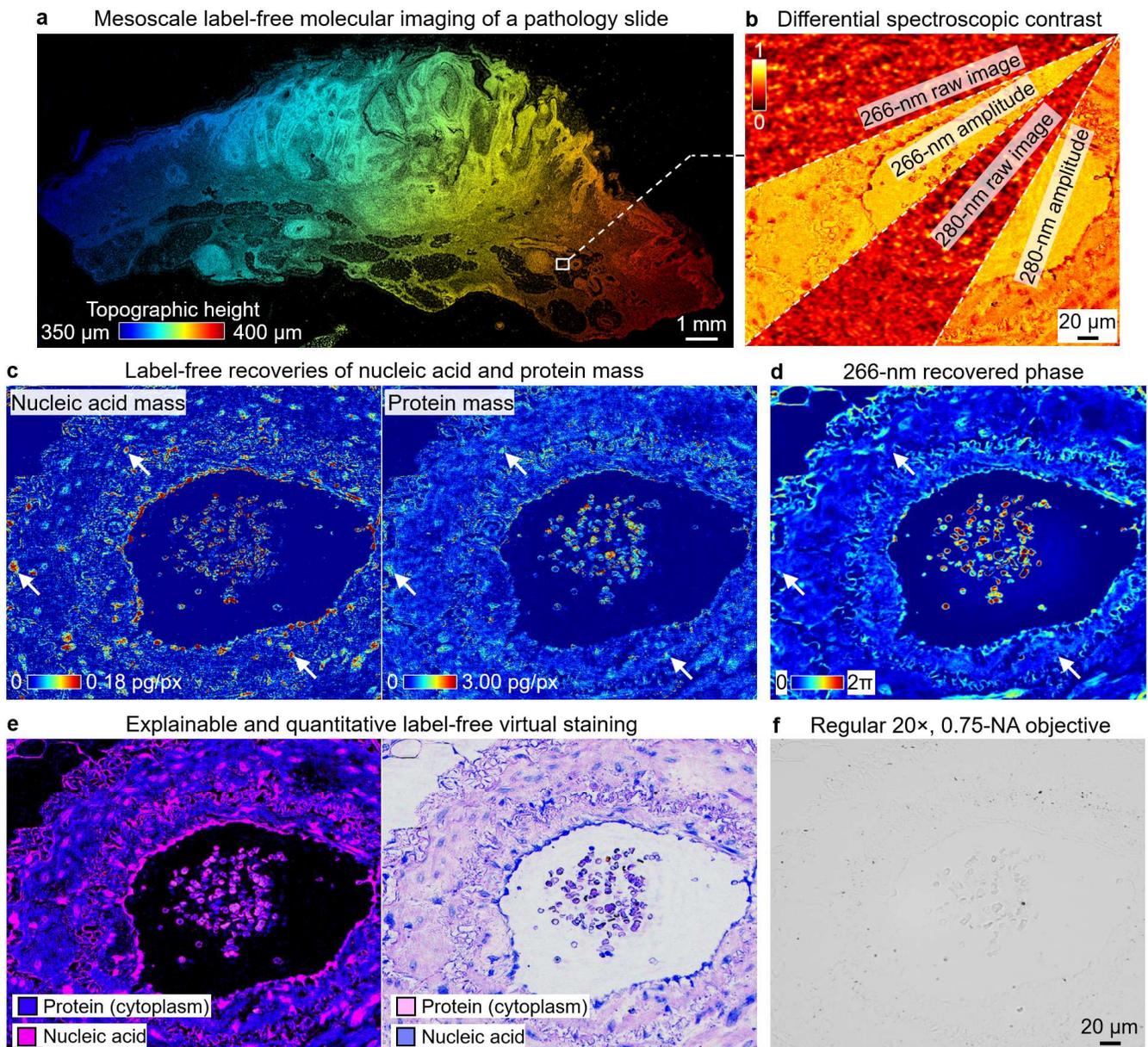

**Fig. 8 | Mesoscale label-free molecular imaging of pathology section using DART. a**, Topographic height map of a label-free pathology slide (5-μm skin tissue section). **b**, Differential spectroscopic contrast images, with raw images and their corresponding recovered amplitudes. **c**, Label-free recoveries of nucleic acid and protein mass distributions. White arrows indicate regions where cell nuclei are identifiable through nucleic acid content. **d**, Recovered phase image of the same sample. Note that the cell nuclei highlighted by white arrows are not easily distinguishable in this quantitative phase image, demonstrating the advantage of DART's molecular contrast over traditional phase imaging techniques. **e**, Explainable and quantitative label-free virtual staining based on the recovered molecular distributions. **f**, Comparative image taken with a regular 20×, 0.75-NA objective.

## 3 Discussions

The DART platform advances label-free mesoscale imaging by uniquely integrating lens-free imaging, portability, and molecular spectroscopy in a single handheld device. By combining DUV spectroscopy with high-throughput ptychography, DART resolves traditional performance trade-offs and enables in-situ, high-resolution molecular imaging across diverse applications[30, 63-69].

DART's differential imaging at 266 nm and 280 nm wavelengths quantitatively maps nucleic acids and proteins, with potential for expansion to additional DUV wavelengths for finer molecular differentiation[16]. Unlike conventional black-box virtual staining methods, which are prone to inaccuracies due to the 'long tail' effect -- where rare or previously unseen tissue types are underrepresented in training data -- DART's physics-based
12

approach ensures consistent performance across diverse samples. This foundation in physical principles provides interpretability crucial for regulatory approval and clinical adoption.

The compact design extends high-resolution imaging beyond traditional laboratories to challenging environments. Potential space applications include monitoring cellular changes in astronauts during extended missions, studying microbial adaptation under microgravity, and analysing extraterrestrial samples for organic signatures. The portable platform could equally serve field diagnostics and resource-limited settings.

Several promising research directions could further enhance DART's capabilities. The virtual error-bin state concept introduced in DART could be extended beyond the perturbed propagation distance. Alternative parameters for creating virtual states could include using slightly perturbed wavelengths or pixel sizes in the forward model, which would similarly create mismatch between the virtual path and the accurate object model. These approaches warrant investigation to determine their relative effectiveness in capturing different types of artifacts. Additionally, incorporating multiple virtual states could potentially further improve reconstruction quality, though determining the optimal number of states and their computational cost would require careful analysis.

Beyond virtual state refinements, incorporating additional DUV wavelengths in a wavelength multiplexed manner[18, 19] could refine molecular differentiation and reduce acquisition time. Implementing the DART concept in a ptycho-endoscopy setting[70, 71] could lead to the development of new portable forensic tools. For instance, a DART-based ptycho-endoscopy could enable high-resolution, label-free imaging of fingerprints or other trace materials using DUV light in hard-to-reach areas, enhancing the detection of trace evidence in crime scenes. Another promising direction is to adopt angle-varied illumination for depth-multiplexed imaging of 3D samples[41, 72, 73] and increase spatial resolution via Fourier aperture synthesis[35, 53]. Combining DART with complementary imaging modalities such as polarimetric[74], fluorescence[75] or ptychographic Raman spectroscopy[76] could provide multi-dimensional information, enhancing its specificity and broadening its application scope. Future research could explore inverse design techniques to optimize surface design computationally. For example, the incorporation of the meta-surface concept[77] may improve performance or introduce additional functionality like detecting polarization properties of the sample.

In the current implementation, we demonstrated label-free molecular imaging in the DUV region. The DART concept can also be extended to the near-infrared region where rich molecular contrast can be obtained through vibration signatures. The computational framework in the current implementation would remain largely applicable, requiring mainly the redesigned differential analysis for specific vibrational features. Potential challenges include inherently lower spatial resolution due to longer wavelengths and weaker vibrational absorption signals compared to DUV electronic transitions. As DART evolves, its potential applications in cytopathologic analysis, in-situ biological monitoring, medical device characterization, and astrobiological research aboard space missions become increasingly feasible, promising to extend mesoscale label-free imaging across multiple disciplines.

## 4 Methods
### 4.1 Experimental setup
The DART setup is a compact, handheld device that integrates DUV spectroscopy with high-throughput ptychography for high-resolution lensless imaging. As detailed in Supplementary Fig. 1 and Note 1, the illumination module consists of three light sources: a 266-nm DUV LED (Crystal IS KL265-50T-SM-WD), a 280-nm DUV LED (Nichia NCSU334A), and a 405-nm laser diode (D405-20, US-Lasers). While DUV laser are commercially available, they are significantly more expensive, require a bulky optical setup to deliver laser beam, with more complex driving electronics and consume more power -- all critical constraints for our pocket-sized platform. DUV LEDs offer a cost-effective solution that supports our goal of creating a widely accessible imaging system while meeting the size and power requirements of a portable device. Since the emitting surface size and shape can be determined with a regular microscope, we can model the LED source as multiple spatially incoherent point source states. This approach enables high-fidelity molecular imaging despite the coherence limitations.

The optical path utilizes UV-enhanced aluminium mirrors (Optical Mirror LLC) with high reflectivity in the DUV region to maximize light efficiency. To enhance the sensor's DUV sensitivity, we modified the image sensor (Sony IMX 226) by removing the protective cover glass and etching away the microlens array using photoresist stripper (a mixture of 99% 1-methyl-2-pyrrolidone, 17% barium sulfate, 11% sulfuric acid, 8% sodium bifluoride, 5% ammonium bifluoride). As demonstrated in Supplementary Fig. S2, these modifications doubled the sensor's sensitivity at DUV wavelengths compared to the original configuration. A crucial component in DART is the coded surface placed on top of the sensor. This surface, fabricated on a thin fused silica substrate, features a disorder-engineered pattern optimized for ptychographic imaging[20].

To enable precise sensor movement, DART incorporates an actuation system inspired by sensor shift technology used in smartphones. As shown in Supplementary Note 1, magnets are coupled voice coil actuators to enable fine control of sensor movement. Importantly, DART recovers positional shifts of the sensor post-measurement, making mechanical motion accuracy not relevant in the reconstruction process. Supplementary Fig. S1 and Note 1 illustrate the integration of ball bearings, which reduce friction and ensure smooth mechanical movement. It shows the coil actuators integrated to control the movement of the sensor and other internal components. Two motion constraint rails guide sample movements in x and y directions, preventing sensor rotation that would complicate the motion tracking process. The entire assembly is housed in a compact enclosure, as seen in Supplementary Fig. S1 and Note 1.

**4.2 DART imaging of label-free samples**
DART enables label-free imaging across different wavelengths to accommodate various biological applications. For molecular contrast imaging using DUV illumination (266 nm and 280 nm), we employed an exposure time of ~0.1 seconds per acquisition to balance signal-to-noise ratio and acquisition speed. A typical dataset consists of ~400 diffraction patterns covering a sensor area of ~40 mm², with a total acquisition time of ~1 minute. Supplementary Fig. S11 demonstrates DART reconstructions with different exposure times at the DUV wavelength. For applications focused on morphological and dynamic features, we utilize 405-nm laser diode illumination. The higher intensity at this wavelength enables significantly faster acquisition, allowing the same area to be imaged in just ~13 seconds during continuous sensor-shift motion.

**4.3 Ptychographic reconstruction with virtual states**
A novelty of the DART reconstruction pipeline is the incorporation of virtual states with intentional perturbations in the forward model. This approach effectively isolates and removes artifacts caused by various error sources, ensuring high-fidelity reconstructions that can resolve different sub-cellular content. In our implementation, we perturb the distance between the code surface and the pixel array to generate the virtual state. The forward imaging model with virtual state correction can be expressed as:

$$I_i(x,y) = \sum_s |OS(x,y)|^2 + \sum_s |VS(x,y)|^2, \quad (1)$$

where $I_i(x,y)$ denotes the $i^{\text{th}}$ captured diffraction pattern. $OS(x,y)$ and $VS(x,y)$ represent the wavefields of the object state and the virtual state, respectively. The subscript 's' refers to different spatially incoherent modes induced by the partially coherent light source. The captured intensity is therefore an incoherent summation of the intensities from the spatially decomposed object state $OS(x,y)$ and virtual state $VS(x,y)$. The wavefields of $OS(x,y)$ and $VS(x,y)$ can be defined as follows:

$$OS(x,y) = \{E_s(x-x_i, y-y_i) \cdot CS(x,y)\} * \text{PSF}_{\text{free}}(d_2) \quad (2)$$

$$VS(x,y) = \{E_{\text{virtual}_s}(x-x_i, y-y_i) \cdot CS_{\text{virtual}}(x,y)\} * \text{PSF}_{\text{free}}(d_2 \cdot a) \quad (3)$$

Here, $E_s(x-x_i, y-y_i)$ and $E_{\text{virtual}_s}(x-x_i, y-y_i)$ represent the shifted diffracted wavefields of the object and the virtual object, respectively, at the plane of the coded surface. The positional shifts $(x_i, y_i)$s are recovered based on the captured image post-measurement[78], making precise tracking unnecessary during acquisition. $CS(x,y)$ and $CS_{\text{virtual}}(x,y)$ denote the coded surfaces corresponding to the object state and virtual error-bin state, respectively. $\text{PSF}_{\text{free}}$ is the point spread function for free space propagation, and $d_2$ represent the distance from the coded surface to the sensor pixels. A perturbation parameter '$a$' in Eq. (3) is introduced to disturb the forward imaging model of DART (in our implementation, parameter '$a$' is 1.1). Supplementary Fig. S12 shows the DART recovered images with values of the parameter '$a$'. The virtual state concept has potential applications beyond light microscopy, offering improvements in electron and X-ray imaging. In electron ptychography, for instance, it could mitigate effects of inelastic scattering and multiple scattering events[79], while in X-ray ptychography, it could address challenges related to partial coherence and other error sources[80, 81].

The reconstruction algorithm is currently implemented in MATLAB and executed on a personal computer (Dell Precision 3680). Due to memory constraints, we typically divide the dataset into 400 images of 512 by 512 pixels each. The reconstruction process involves an iterative algorithm that alternates between updating object and virtual object wavefields and enforcing consistency with the measured diffraction patterns (Supplementary Note 2). We implement a multi-resolution approach, starting with a low-resolution reconstruction and progressively increasing resolution. For the 405-nm laser diode, where spatial mode decomposition is unnecessary, reconstruction time reduces to ~40 seconds for the same size of dataset. For DUV LED sources requiring spatial mode decomposition, the reconstruction time for 400 diffraction patterns is ~5 minutes, with 24 seconds for the low-resolution reconstruction followed by high-resolution refinements. We note that the current implementation is not optimized for speed. Parallel processing of multiple dataset patches can be implemented on multi-core systems. Optimized routines using CUDA (compute unified device architecture) can substantially accelerate computations. Momentum acceleration techniques and advanced optimization algorithms could also be employed to speed up convergence. Recent advances in deep learning approaches show particular promise, with implicit neural representations, physics-informed neural networks, and direct end-to-end reconstruction networks all demonstrating potential for accelerating ptychographic reconstruction while maintaining fidelity[82, 83].

### 4.4 Nucleic acid and protein mass recovery

Based on reconstructions at 266 and 280 nm, we performed them in the spatial domain via an image registration algorithm[84]. The masses of proteins and nucleic acids were calculated on a per-pixel basis by the Beer-Lambert law:

$$I_{sample}(x,y) = I_{background}(x,y) 10^{-\varepsilon l c} \tag{4}$$

where $I_{sample}$ represents the intensity of DUV light absorbed by the sample, and $I_{background}$ denotes the intensity of deep ultraviolet light transmitted through an empty region, which is used to define the transmission. The variable $\varepsilon$ is the decadic molar extinction coefficient, $c$ indicates the concentration, and $l$ denotes the path length, collectively describing the attenuation properties of the sample. Solving this equation reveals the relationship between optical density and concentration:

$$\varepsilon l c(x,y) = \log\left(I_{background}(x,y)\right) - \log\left(I_{sample}(x,y)\right) \tag{5}$$

Since both proteins and nucleic acids contribute to the measured optical density (OD) in direct proportion to their respective concentrations. The $OD_\lambda(x,y)$ at a given wavelength $\lambda$ can be expressed as a function of the combined concentrations of proteins and nucleic acids:

$$OD_\lambda(x,y) = \varepsilon_\lambda^{pro} \cdot c^{pro}(x,y) \cdot l(x,y) + \varepsilon_\lambda^{nuc} \cdot c^{nuc}(x,y) \cdot l(x,y) \tag{6}$$

The superscripts 'pro' and 'nuc' denote proteins and nucleic acids, respectively. After obtaining the $OD_\lambda(x,y)$ at different wavelengths, the amount of proteins $n^{pro}(x,y)$ and nucleic acids $n^{nuc}(x,y)$ per unit area can be expressed as a function of their respective optical densities at those wavelengths:

$$n^{pro}(x,y) = \frac{OD_{266}(x,y) \cdot \varepsilon_{280}^{nuc} - OD_{280}(x,y) \cdot \varepsilon_{266}^{nuc}}{\varepsilon_{266}^{pro} \cdot \varepsilon_{280}^{nuc} - \varepsilon_{280}^{pro} \cdot \varepsilon_{266}^{nuc}} \tag{7}$$

$$n^{nuc}(x,y) = \frac{OD_{266}(x,y) \cdot \varepsilon_{280}^{pro} - OD_{280}(x,y) \cdot \varepsilon_{266}^{pro}}{\varepsilon_{280}^{pro} \cdot \varepsilon_{266}^{nuc} - \varepsilon_{266}^{pro} \cdot \varepsilon_{280}^{nuc}} \tag{8}$$

By subsequently multiplying the $n^{pro}(x,y)$ and $n^{nuc}(x,y)$ by the area of each pixel and by the molar mass of proteins, $mass_{pro}$, and nucleic acids, $mass_{nuc}$, the final mass maps, $m^{pro}(x,y)$ and $m^{nuc}(x,y)$, are generated for Figs. 7b, 7f, and 8c:

$$m^{pro}(x,y) = n^{pro}(x,y) \cdot (pixelsize)^2 \cdot mass_{pro} \tag{9}$$
$$m^{nuc}(x,y) = n^{nuc}(x,y) \cdot (pixelsize)^2 \cdot mass_{nuc} \tag{10}$$

### 4.5 Quantitative virtual staining

Our virtual staining process is based directly on the recovered nucleic acid and protein mass distributions obtained through in Eqs. (9)-(10). This approach ensures that the virtual stains are grounded in actual molecular content rather than relying on non-specific dye binding or machine learning algorithms. We developed two main types of virtual staining. The first is virtual fluorescence imaging. In this mode, we generate images that mimic fluorescence microscopy. Nucleic acid mass is represented by blue color tones, while protein mass is depicted in purple. The intensity of each color directly correlates with the quantity of the respective biomolecule. The second type is H&E-like staining. In this case, nucleic acid mass is represented by blue color tones, analogous to hematoxylin staining of cell nuclei in traditional H&E. Protein mass is depicted in pink, similar to the eosin staining of cytoplasmic proteins. To create the white background of H&E images, we add a constant value as green color across the image.

**Acknowledgements**
This work was partially supported by the National Institute of Health R01-EB034744 (G. Z.), the UConn SPARK grant (G. Z.), National Science Foundation 2012140 (G. Z.), Department of Energy SC0025582 (G. Z.), National Institute of Health R35-GM147437 (F. E. R.), and National Science Foundation CAREER Award 1752011 (F. E. R.). Q. Z. acknowledges the support of the UConn GE Fellowship. We thank Linnaea Ostroff and Jario Orea from the Department of Physiology and Neurobiology at UConn for preparing the unstained brain samples.

**Competing interests**
G. Z. is an inventor of intellectual property related to the reported system. The other authors declare no competing interests.

**Availability of data and materials**
The data that support the findings are available from the corresponding author on reasonable request.

**Author contributions**
G. Z. conceived the concept of DART and supervised the project. R. W. and Q. Z. developed the prototype systems and performed the imaging experiments. R. W., Q. H., T. W., P. S., and G. Z. prepared the display items. J. Q. and M. M. prepared the de-identified cytopathological samples. R. W. and Q. Z. prepared the Supplementary Notes. All authors contributed to the writing and revision of the manuscript.